\documentclass[sn-basic,Numbered]{sn-jnl}

\usepackage{graphicx}%
\usepackage{multirow}%
\usepackage{amsmath,amssymb,amsfonts}%
\usepackage{amsmath}
\usepackage{amsthm}%
\usepackage{mathrsfs}%
\usepackage[title]{appendix}%
\usepackage{xcolor}%
\usepackage{textcomp}%
\usepackage{manyfoot}%
\usepackage{booktabs}%
\usepackage{algorithm}%
\usepackage{algorithmicx}%
\usepackage{algpseudocode}%
\usepackage{listings}%
\usepackage{geometry}%
\theoremstyle{thmstyleone}%
\newtheorem{theorem}{Theorem}
\theoremstyle{thmstyletwo}%

\theoremstyle{thmstylethree}%

\begin{document}

\title[Article Title]{Modeling and indexing the drought severity by augmenting SPI metric system to incorporate multi-modal ground temperature data}


\author*[1]{\fnm{Sachini} \sur{Karunarathne}}\email{sachinik.maths@stu.cmb.ac.lk}

\author[2]{\fnm{Kushani} \sur{De Silva}}\email{kdesilva@maths.cmb.ac.lk}
\equalcont{These authors contributed equally to this work.}

\author[3]{\fnm{Sanjeewa} \sur{Perera}}\email{ssnp@maths.cmb.ac.lk}
\equalcont{Sachini Karunarathne: methodology and writing, Kushani De Silva: conceptualization, methodology, writing/editing, Sanjeewa Perera: conceptualization}

\affil*[1,2,3]{\orgdiv{Department of Mathematics}, \orgname{University of Colombo}, \orgaddress{\city{Colombo}, \country{Sri Lanka}}}


\abstract{ Drought is a global threat caused by the persistent challenges of climate change. It is important to identify drought conditions based on the weather variables and their patterns.  In this study, we enhanced the Standardized Precipitation Index (SPI) by integrating ground temperature data to develop a more comprehensive metric for evaluating drought severity. Our metric offers a dual assessment of drought severity, taking into account both the intensity of the drought and its duration. We employ this evaluation in the primary paddy cultivation region of Sri Lanka, with the aim of shedding light on the prevailing drought conditions affecting paddy crops due to insufficient water supply and prolonged periods of elevated temperatures. Additionally, we calibrate our metric by aligning it with historical drought records and subsequently compare the outcomes with those derived from the conventional SPI index.}

\keywords{Drought Severity, SPI, MSDI, Paddy, Copula, Multi-modal Distributions}



\maketitle

\section{Introduction}\label{sec1}
Droughts represent a substantial threat to nations and their inhabitants in terms of economy, food security, health and much more. Currently, multiple regions worldwide are suffering with droughts at varying degrees of severity. The examination of droughts is of utmost importance, considering that the likelihood of these events becoming more intense in the future is high, primarily due to the persistent challenges posed by climate change. Concerning food security and its relationship with climate change, rice, plays a pivotal role. Rice serves as a staple food for more than half of the global population \cite{fukagawa2019rice}, particularly in Asia, where it serves as a fundamental source of calories and nutrition. Moreover, paddy fields contribute to environmental sustainability by acting as carbon sinks, aiding in the mitigation of climate change. In particular, paddy cultivation is of paramount importance to Sri Lanka, both historically and in the present day. The country's culture, economy, social fabric and food security have deep ties to rice production \cite{ref5,ref6}. 
North Central province (NC) of Sri Lanka makes the largest contribution to the country's agriculture and enable 8.1\% of country's GDP and approximately 107 kg per person of daily consumption necessity for Sri Lankan population \cite{ref9,ref10,ref11,ref12,ref13,ref14}. Paddy cultivation is supported by the weather of two main monsoons  referred to as Yala and Maha - Maha seasons from September to March and Yala season from May to August. NC experiences a combination of both of these weather patterns at a given month due to its geographical location.\\
Analyzing individual factors contributing to droughts falls short in offering a comprehensive understanding of the complex ramifications of drought events \cite{ref31,ref32}. Consequently, relying solely on metrics like Standardized Precipitation Index (SPI) proves insufficient in gauging drought severity because it primarily assesses the influence of rainfall \cite{ref30,ref48}. However, temperature is a crucial parameter in discerning droughts since certain temperature conditions can offset the effects of rainfall. Moreover, the conventional SPI metric is based on the assumption that rainfall adheres to a gamma distribution \cite{ref49}. However, our research has unveiled that the gamma distribution is insufficient for effectively modeling rainfall data. As a result, we cast doubt on the appropriateness of employing this distribution for the calculation of any drought index. This matter becomes particularly pertinent in regions where meteorological variables do not conform to a single distribution but exhibit multiple distributions due to fluctuating seasonal weather patterns. We have encountered such a situation in the NC province, where the influence of two distinct monsoon seasons has caused temperature data to deviate from a unified probability distribution framework. \\
It is understood that hazards such as droughts are caused by extreme weather conditions where conventional joint distributions would fail to model. Therefore, it is important to choose a proper model that enables to model extreme cases of weather that would pay more attention to the tails of the respective marginal distributions. Copula is one of the robust frameworks that enable modeling tail dependencies, and by that, it allows to capture the consequences of extreme weathers that lie in tails of the distributions \cite{ref34}. The selection of a suitable copula is based upon meticulous considerations such as Kendall's tau rank correlation measure and the shapes of the marginal distributions of the selected variables \cite{ref26}. The Kendall's tau $\left( \tau\right) $, measures the rank correlation providing insights into the strength and direction of dependence between variables. Several examples of modeling extreme cases of drought variables can be found at \cite{otkur2021copula,seyedabadi2020multivariate} and references therein. However, in these studies, variables such as drought severity and drought duration have been used as the drought variables \cite{ref26,ref41,ref42}. In contrast we build the index by directly using the weather data avoiding the uncertainty in calculated drought variables.\\
In this study, we introduce an enhanced calculation for SPI that provides a more precise representation of rainfall by utilizing a more fitting probability distribution. Furthermore, we incorporate temperature data to construct a more accurate metric for measuring drought. This involves addressing the challenge of modeling ground temperature in regions characterized by multiple weather seasons. Subsequently, we establish a drought index capable of assessing drought severity from two perspectives, and we compare our findings with those obtained using the traditional SPI index. Our new metric's validity is verified through an examination of historical drought records, offering valuable insights into paddy-related drought conditions.\\
The structure of this paper is as follows: Section 2 outlines the methodology for fitting an appropriate copula to the marginal probability distributions of selected weather variables, while Section 3 details the development of the drought measuring metric and the process of  its calibration followed by Results and Discussion. The paper concludes with a summary in the final section with valuable future avenues to explore.


\section{Jointly dependent weather model}

Driven by the rationale presented in the introduction, we have chosen NC as our study area (latitude: $8.1996$ \textdegree{}N and longitude: $80.6327$ \textdegree{}E), covering a span of 40 years from 1981 to 2021 to demonstrate the applicability of the contributions of this work. The average rainfall and temperature conditions in NC and favorable thresholds of these variables for the paddy are given in Table \ref{tab:NCP-details}. According to Table \ref{tab:NCP-details}, although the favorable conditions of rainfall and temperature agrees with the district wise monthly average values, NC has experienced drought conditions numerous times in the past \cite{ref9}. This observation underscores the need for multivariate modeling, emphasizing the importance of not relying solely on SPI and advocating for the inclusion of prolonged periods in drought decision-making processes \cite{ref43}. To achieve these goals, we model the drought conditions using copula functions that enable identifying joint structures between variables. The monthly average rainfall (mm/day) and monthly temperature data (\textdegree C) were retrieved from the NASA website \cite{NASApower}. In the next subsections, we demonstrate how the marginal probability distributions are fitted to rainfall and temperature data respectively to finally showcase the joint model using copula.


\begin{table}[ht]
	\centering
	\caption{Weather characteristics of the NC province's districts are illustrated \cite{ref21}. The favorable environmental conditions are also given for comparisons \cite{ref22}. Although the average rainfall and temperature values lie in the favorable bounds, NC has faced droughts for paddy numerous times. The annual favorable values are not shown as they do not make any useful contribution to the study.}
	\begin{tabular}{l p{1.8cm} p{1.7cm} p{1.5cm} p{1.5cm} }
		\toprule[1.5pt]
		District & Monthly Average Rainfall (mm/day) & Monthly Average Temperature (\textdegree{}C) &Total Annual Rainfall & Avg. Annual Temperature  \\
		\midrule
		Anuradhapura & $3.6$  & $28.3$& 1533.4 mm & 28.2 \textdegree{} C \\
		Polonnaruwa & $3.2$ & $31.0$ & 1698.3 mm & 28.8 \textdegree{} C \\
		\midrule
		NC province & $4.3$ & $28.9$ & $1589.0$ mm & $27.3$ \textdegree{} C \\
		\midrule
		Favorable conditions & $\left(2.74 - 4.11\right)$ & $\left(21 - 27\right) $ & &\\
		\bottomrule
	\end{tabular}
	\label{tab:NCP-details}
\end{table}

\begin{figure}[ht]
	\centering
	\includegraphics[width=1\textwidth]{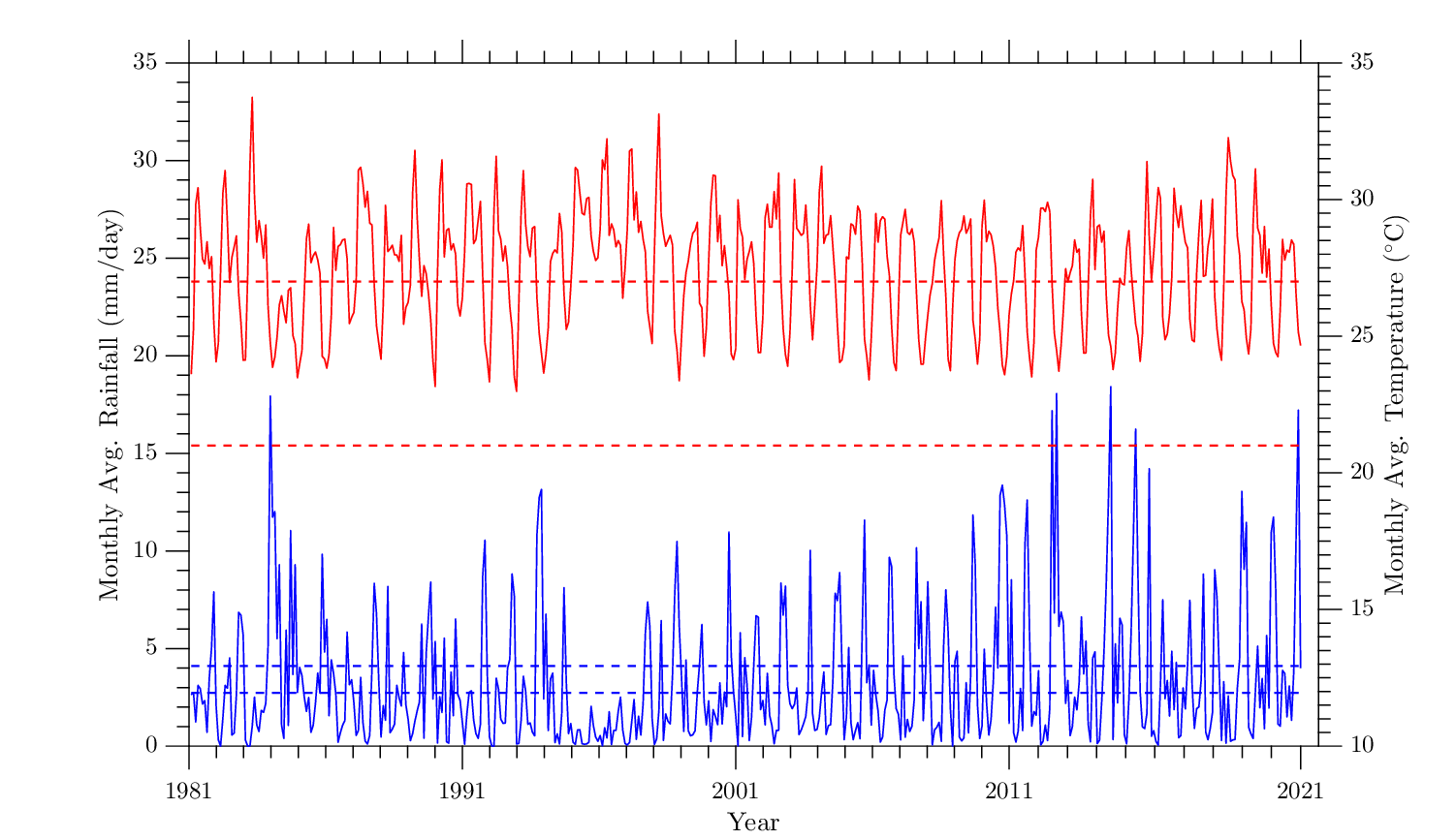}
	\caption{Time series plots of monthly average temperature and rainfall in NC for the period of 1981-2021. The favorable thresholds for paddy are marked with dashed horizontal lines.} 
	\label{T}
\end{figure}



\subsection{Estimating the marginal distributions}

\begin{figure}[ht]
	\centering
	\includegraphics[width=1.1\textwidth]{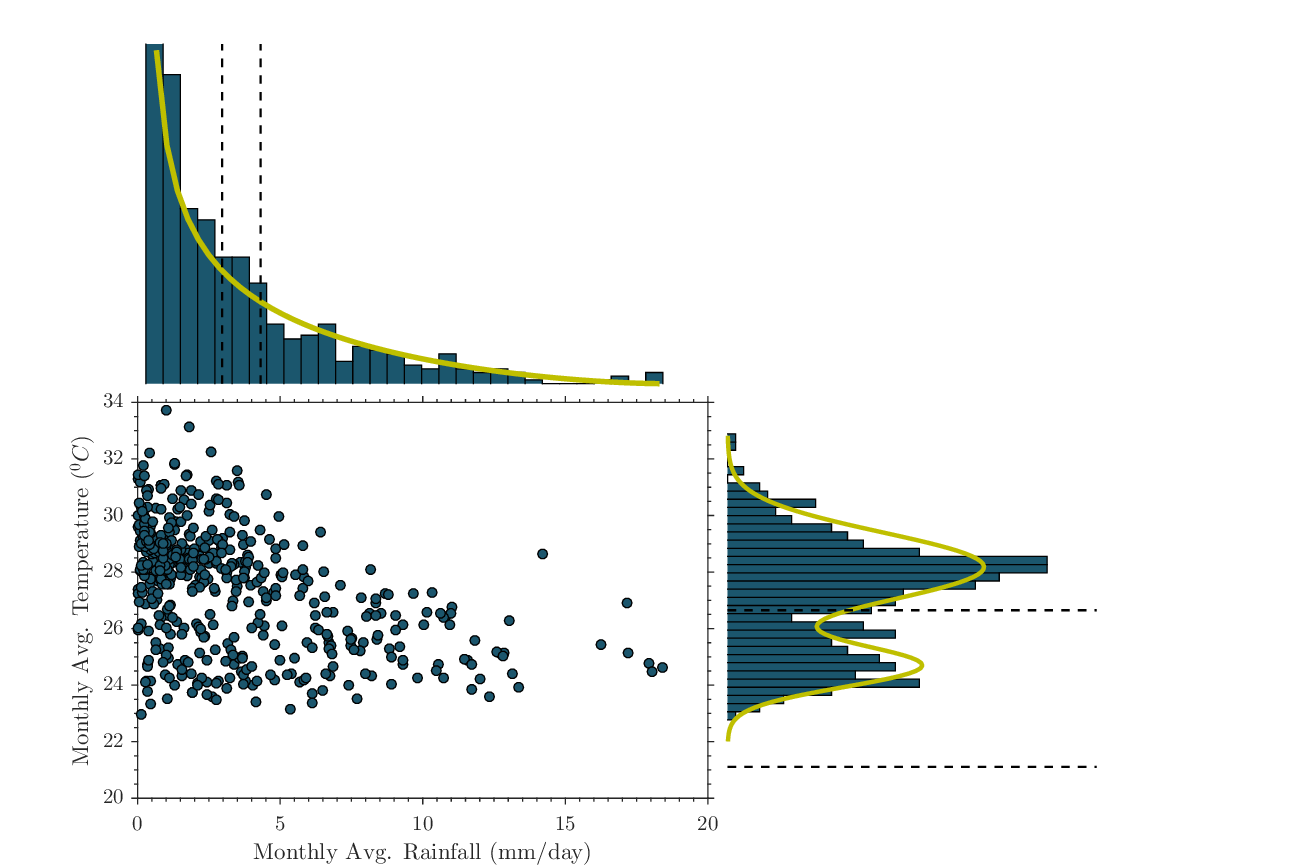}
	\caption{Histogram of monthly average rainfall and temperature data. Rainfall data are rightly skewed distributed in the range of $\left( 0-18.41\right) $ mm/day. The yellow line shows the best-fitted beta distribution for rainfall. Histogram of monthly average temperature shows a multi-modal behavior where a local peak within the favorable temperatures (vertical dashed lines) and the global peak is outside the favorable temperature. The data are distributed in the range of (22.98-33.74) \textdegree C. The yellow solid line shows the GMM fitted to the temperature data.}
	\label{fig:marginals}
\end{figure}

In order to determine the marginal probability distributions of each variable, the rainfall and temperature data were fitted to different known probability functions. Afterwards, the best fit was chosen based on AIC criteria $	\text{AIC} = 2k -2\ln({\mathcal{L}}),$ where $k$ and ${\mathcal{L}}$ are respectively the number of parameters and the likelihood function of the model \cite{ref25}. A likelihood function defined on the parameter space $\Omega$ is given by,
\begin{equation}\label{eq:like}
	\mathcal{L}(\Theta \mid x_1, x_2, \ldots, x_n) = \prod_{i=1}^{n} f(x_i; \Theta)
\end{equation}
where $\Theta$ is the vector of model parameters, $x_i,i=1,\cdots,n$ represent the $n-$ sized dataset and $f(.)$ is the probability density function of $x_i$. According to the AIC, the rainfall data (Fig. \ref{fig:marginals}) agrees the beta distribution rather than the gamma distribution (see Table \ref{tab:univariate_rainfall}). This is evident that gamma distribution framework is lacking to represent rainfall. Thus, this measurement has increased the accuracy of a conventional SPI value. The mean of rainfall obtained from the fitted beta distribution is $E(R)=0.128$ and it is outside of the favorable rainfall for paddy. Moreover, the probability of NC rainfall (R) being favorable to paddy is $14.04$. i.e. $p\left(R\in \left( 2.74 \text{ mm },4.11 \text{ mm }\right)\right) = 14.04\%$ which indicates that 86\% of the time, rainfall in NC is not favorable to paddy.
\begin{table}[ht]
	\caption{Rainfall: fitted probability distributions and their corresponding AIC values for rainfall data. HG represents the Half-Gaussian distribution, i.e. $N(\mu,\sigma)$ where $\mu=0$ and $\sigma \in \left( 0,\infty\right) $. }
	\label{tab:univariate_rainfall}
	\centering
	\begin{tabular}{ lll }
		\toprule[1.5pt]
		Probability distribution & Parameter estimates & AIC value \\
		\midrule
		Beta $(\alpha_{R},\beta_{R})$ &$\alpha_{R}=0.779$, $\beta_{R}=5.271$ & 2061.749\\
		Gamma $(k_{R},\theta_{R})$ &$k_{R}=0.861$, $\theta_{R}= 3.690$ & 2104.167\\
		HG $(\mu_{R},\sigma_{R})$ & $\mu_{R}= 0, \sigma_{R}= 4.821$ & 2265.897\\
		\bottomrule
	\end{tabular}
\end{table}
\\
The histogram of temperature data depicted in Fig. \ref{fig:marginals} shows clear evidence of non-Gaussian multi-modals. The two peaks of the histogram fairly agrees to that of a Gaussian, i.e. assuming data are generated by a combination of two Gaussian distributions. The Gaussian Mixed Model (GMM) of two components is fitted to the temperature data and it is mathematically denoted as below:
\begin{equation}
	f(x|\Theta_i ) = \sum_{i=1}^{n} w_i \cdot \mathcal{N}(\mu_i, \sigma_i^2),
	\label{gmm1_eq}
\end{equation}
where $\Theta_i=\left( w_i,\mu_i,\sigma_i\right)$. Here, $w_i, \sum_{i}w_i=1$ denotes the weight assigned to each component distributed with mean $\mu_i$ and $\sigma_i$. The parameters were estimated using the iterative Expectation-Maximization (EM) algorithm where initial guesses for the parameters were obtained by K-clustering algorithm. The parameter estimates for $\Theta_i = \left(w_i,\mu_i,\sigma_i \right) $ were estimated for two components respectively as $\Theta_1 = \left( 0.31,24.8586,0.8015\right) $ and $\Theta_2 = \left(0.69 , 28.6589, 1.3191\right) $. The final fitted density is given in Fig. \ref{fig:marginals}.

\subsection{Copula fitting}
A copula is a function that joints several one-dimensional distribution functions to a multivariate distribution function by capturing their dependency structure \cite{ref23}. For a bivarite case with variables $X$, $Y$ and marginal distribution functions $F_X(x)$, $F_Y(y)$ then there exists a copula $C$ such that,
\begin{equation}
	\label{copula}
	P(X \le x,Y \le y)=C(F_X(x),F_Y(y)).
\end{equation}
The copula $C$ is fitted to the transformed uniform data ($U_X$ and $U_Y$) of the marginals. PIT (see Theorem \ref{thm:PIT}) gurantees the uniformity of the random variables found by transforming marginal distribution functions $F_X(x)$, $F_Y(y)$ \cite{ref24}.\\

\begin{theorem} \label{thm:PIT}
	Probability Integral Transform Theorem (PIT) \\
	If \(X\) is a continuous random variable with distribution function \(F\), then the random variable \(U = F(X)\) (the probability integral transform of \(X\)) is uniformly distributed on the unit interval \(I = [0,1]\); or equivalently, \(P[F(X) \leq t] = t\) for all $t \in I$.\\ 
\end{theorem}

\begin{figure}[ht]
	\centering
	\includegraphics[width=1\textwidth]{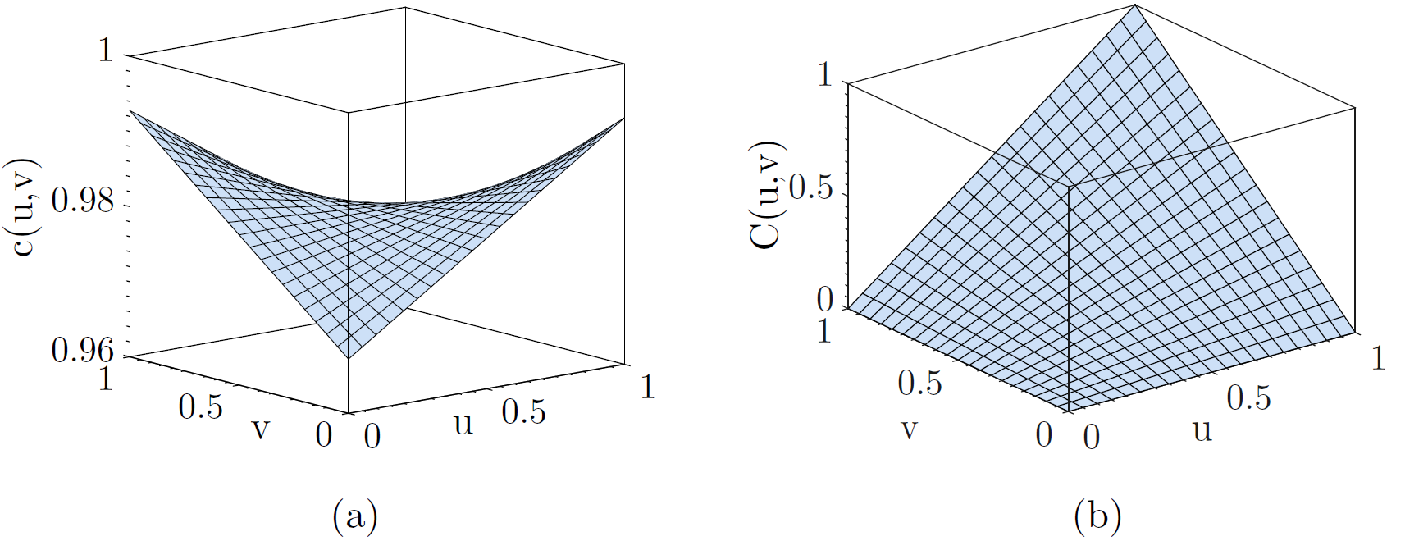}
	\caption{ 
		(a) Fitted FGM copula density and (b) Fitted FGM copula distribution at $\theta= -0.0319$ . The axis labels $u,v$ respectively show the transformed rainfall and temperature data.} 
	\label{fig:fgm_density}
\end{figure}

The marginal densities obtained in the previous section is then transformed into uniform distributions using PIT to identify a suitable copula function. Obtaining a best-fit copula is done in two steps: (a) finding the appropriate copula from set of pre-defined copula functions using a goodness of fit test, (b) optimizing the tune-in parameter of the selected copula. 
The value of Kendall's $\tau \in \left[ -1,1\right]$ was $ -0.00711$ for the uniform data of rainfall ($u$) and temperature ($v$). Based on this value of $\tau$, Frank and Farlie-Gumbel Morgenstern (FGM) copulas were suitable as they are capable of capturing low dependencies. The selected copulas, their distributions and density functions are given in Table \ref{tab:1} \cite{ref27,ref26}. To find the proper dependency structure between weather axes, these two copulas were fitted to uniform data. 
\begin{table}
	\centering
	\caption{Set of appropriate copula functions according to the resulting low dependency Kendall's tau rank correlation. The distribution function and density function of each copula is given by uppercase and lowercase $c(.)$ respectively. The copula parameter is represented by $\theta$ and uniform data for rainfall and temperature are given by $u,v$ respectively. The estimated copula parameter and the p-value of the goodness of fit values ($S_n$) are also given.}
	\label{tab:1}
	\begin{tabular}{lll}
		\toprule[1.5pt]
		& Frank  &  Farlie-Gumbel-Morgenstern (FGM) \\
		\midrule
		$C(u,v)$ & $\dfrac{1}{\theta} \left(1+\dfrac{(e^{-\theta u} - 1)(e^{-\theta v} - 1)}{e^{-\theta}-1}\right)$ & $uv\left( 1 + \theta (1 - u)(1 - v)\right) $\\
		\midrule
		$c(u,v)$ & $ -\dfrac{\theta e^{-\theta (u+v)} (e^{-\theta}-1)}{(e^{-\theta (u+v)} - e^{-\theta u} -e^{-\theta v} + e^{-\theta})^2}$ & $1 + \theta(1 - 2u)(1 - 2v)$ \\
		\midrule
		Estimated $\theta$ & $-0.0013$ & $-0.03195$\\
		\midrule
		p-value of $S_n$ & $0.6389$ & $0.6968$\\
		\bottomrule
	\end{tabular}
\end{table}The most appropriate copula is selected by following the hypothesis testing given below:
\begin{description}
	\item [$H_0$]: The given copula $C_{0}$ belongs to the chosen parametric copula family $C$.
	\item [$H_1$]: The given copula $C_{0}$ does not belong to the chosen parametric copula family $C$.
\end{description}
The test-statistic to select the correct hypothesis is the Cramer-von Mises statistic $S_{n}$ for $n$ number of data points \cite{ref28,ref17},
\begin{equation}
	S_n = \sum_{i=1}^{n} (C_n({u}_{in}, {v}_{in}) - C_{\theta_n}({u}_{in}, {v}_{in}))^2, \quad 
	\label{eq:sn}
\end{equation}
where $C_{n}=\frac{1}{n}\sum_{i=1}^{n}1 (u_{in} \leq u, v_{in} \leq v)$, and $u_{in}, v_{in}$ are given by $\frac{1}{n+1}\sum_{j=1}^{n}1 (X_{j} \leq X_{i})$ and $\frac{1}{n+1}\sum_{j=1}^{n}1 (Y_{j} \leq Y_{i})$ respectively. The empirical copula $C_{n}$ was obtained from the pseudo-observations $\left( u_{in},v_{in}\right)_{i=1}^n$ calculated from $(X_{1}, Y_{1}),...(X_{n}, Y_{n})$ where $X$ and $Y$ are the rainfall and temperature data. 
According to Table \ref{tab:1}, the p-values of both copula are significant at 5\% level. The strongest copula is therefore selected to be FGM as it has the highest p-value \cite{ref17}. The FGM copula density plot and contour plot is given in Fig. \ref{fig:fgm_density}.

\section{Developing the drought severity metric }
In this section we showcase how the copula model is used to calculate the severity metric. The joint distribution with cumulative joint probability $p$ for weather data can be expressed with the FGM copula (see Eq. \eqref{cop}).
\begin{equation}
	P(X \leq x, Y \leq y) =C[F(X),G(Y)]=p,
	\label{cop}
\end{equation}
where $F(X)$ and $G(Y)$ are the marginal cumulative distribution functions of rainfall ($X$) and temperature ($Y$). In \cite{ref30}, they have introduced a method of calculating their MSDI metric using, 
\begin{equation}
	MSDI =\psi^{-1}(p),
	\label{bsdi}
\end{equation}
where $\psi$ is the standard normal distribution function. We simply borrow Eq. \eqref{bsdi} to calculate MSDI for our own variables. The metric calculated from Eq. \eqref{bsdi} gurantees to generate values in the range of $\left( -\infty,\infty\right) $ by the defined range of standard normal distribution. The infinity for the index is approximately taken as 5 due to the fact that 99\% of the probability of the normal distribution is between $\left(-4,4 \right) $. Negative and positive MSDI values respectively indicate dry climate conditions (drought) and wet climate conditions. MSDI near zero signify the normal climate conditions \cite{ref30}. The indices generated in the previous section were then re-computed for different time periods to showcase the prolonged period of certain weather dependencies. Thus the two dimensions of the drought assessment system are MSDI index and the prolonged period. Four significant time periods were identified for T as 3-month, 6-month, 9-month and 12-month. When the prolonged period, T is higher, the dry conditions from MSDI reduced because less dry conditions for a longer period can cause for severe droughts. The n-month MSDI is computed as,
\begin{equation}
	n\text{-month MSDI} = \frac{1}{n} \sum_{k=1}^{n} \text{1-month MSDI}_k.
	\label{msdi}
\end{equation}

\subsection{System calibration}
\begin{figure}
	\centering
	\includegraphics[width=0.8\textwidth]{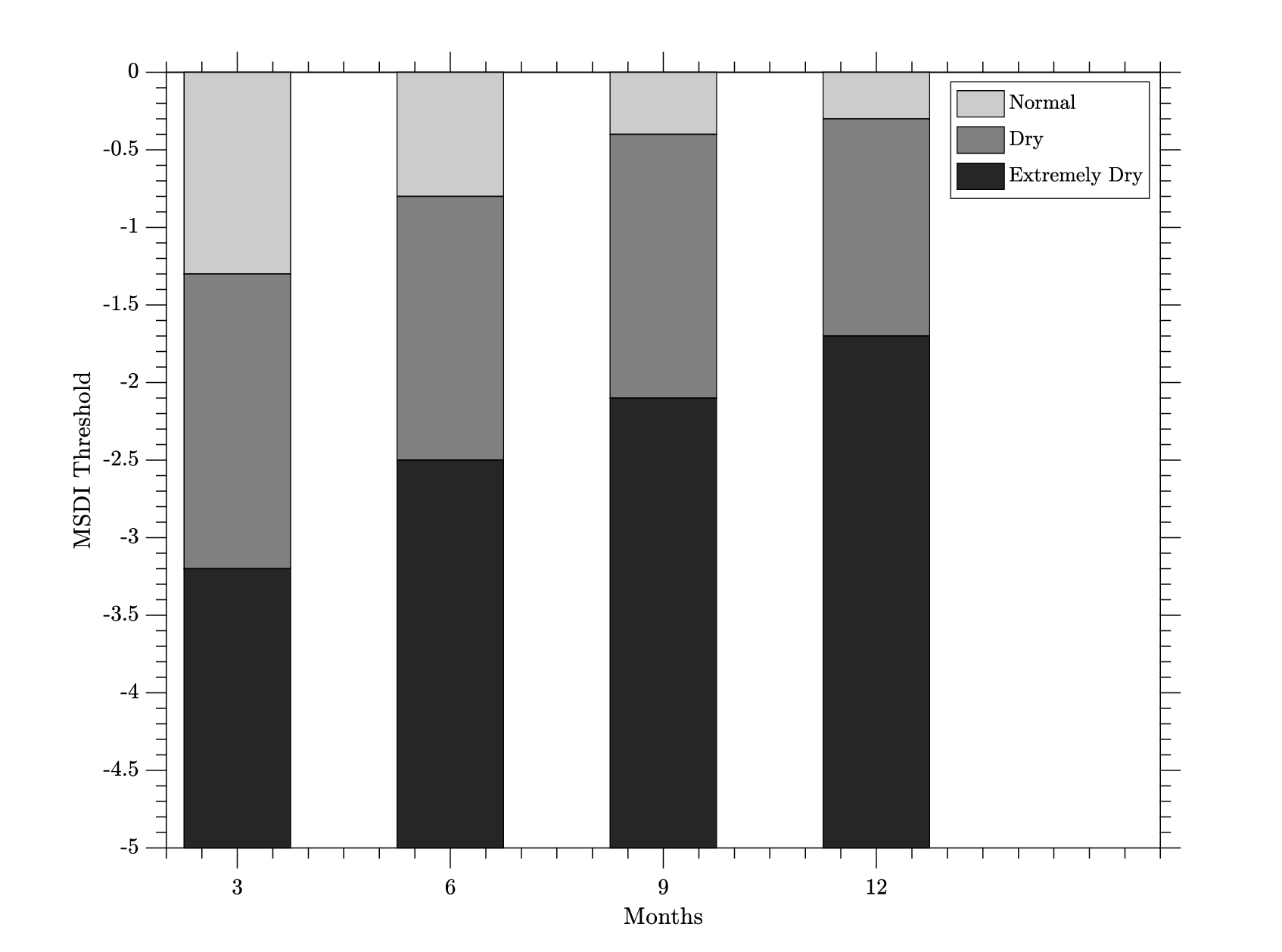}
	\caption{The trend of thresholds of normal, dry and extremely dry conditions along with prolonged period is depicted. } 
	\label{fig: MSDI limits}
\end{figure}

The MSDI values generated for different T periods are categorized based on identified thresholds to identify normal conditions and two different dry conditions - namely dry and extremely dry. The thresholds identified can be found at Table \ref{tab:tresh}. For the two dry conditions, the thresholds of MSDI in different T periods are depicted in Fig. \ref{fig: MSDI limits} for more clarity. According to the Fig. \ref{fig: MSDI limits}, when the prolonged period, T increases, the severity of dry weather conditions decreases indicating severity $\propto \left( T \times MSDI\right) $. Moreover, when comparing the dry conditions vs extremely dry conditions, the levels of thresholds in MSDI are smaller for extremely dry conditions. The threshold to differentiate dry from extremely dry increases as T increases. In summary these trends indicate that we can reliably identify a drought situation when there is a combination of high temperatures and low rainfall persisting over an extended period. However, it's important to recognize that when we examine shorter time spans, it becomes more challenging to swiftly determine whether a drought condition is present or not. Therefore, the duration of the observed climate conditions significantly influences our ability to classify them as a drought. 

\begin{table}
	\centering
	\caption{Two-dimensional threshold values for MSDI to identify drought conditions are given. }
	\label{tab:tresh}
	\begin{tabular}{lcccc}
		\toprule[1.5pt]
		\multicolumn{1}{c}{\multirow{2}{*}{\textbf{Drought conditions}}} & \multicolumn{4}{c}{\textbf{MSDI range}} \\
		\cmidrule{2-5}
		& \textbf{3-month} & \textbf{6-month} & \textbf{9-month} & \textbf{12-month} \\
		\midrule
		Extremely dry & $\left( -\infty,-3.2\right) $ & $\left( -\infty,-2.5\right) $ & $\left( -\infty,-2.1\right) $ & $\left( -\infty,-1.7\right) $ \\
		Dry & $\left( -3.2 , -1.3\right) $ & $\left(-2.5,-0.8\right) $ & $\left(-2.1,-0.4\right) $ & $\left(-1.7,-0.3\right) $ \\
		Normal & $\left( -1.3 , +\infty\right) $ & $\left( -0.8 , +\infty\right) $ & $\left( -0.4 , +\infty\right) $ & $\left( -0.3 , +\infty\right) $ \\
		\bottomrule
	\end{tabular}
\end{table}

\section{Results and discussion}
\begin{figure}
	\centering
	\includegraphics[width=1\textwidth]{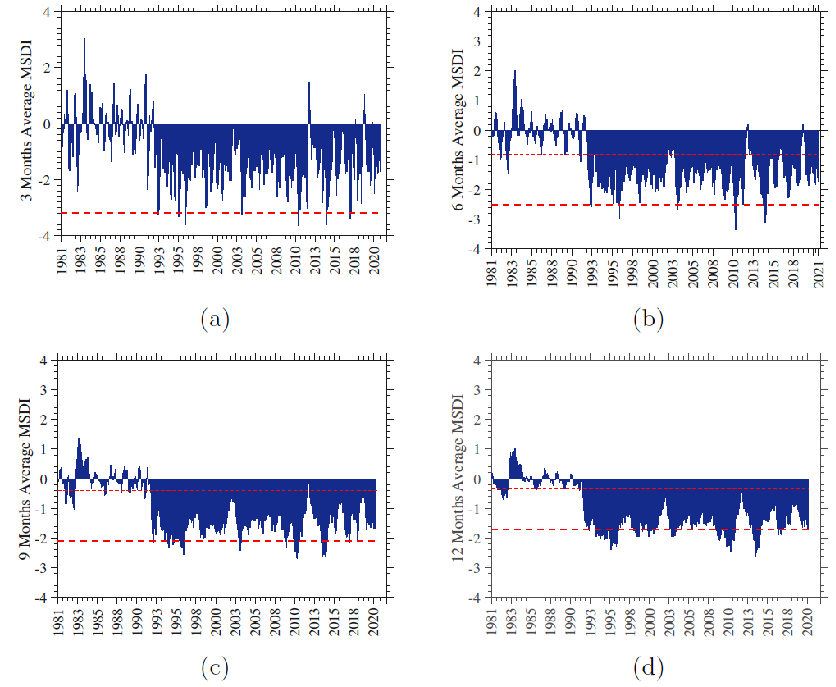}
	\caption{The MSDI for 3, 6, 9, 12 months for the period of 1981-2021 are shown respectively in (a), (b), (c) and (d). The highest downward peaks represent the years of the occurrence of severe droughts. The red(bold) and red(thin) dash lines represent threshold limits for extremely dry and dry conditions respectively.} 
	\label{MSDI_panel}
\end{figure}

\begin{figure}
	\centering
	\includegraphics[width=1\textwidth]{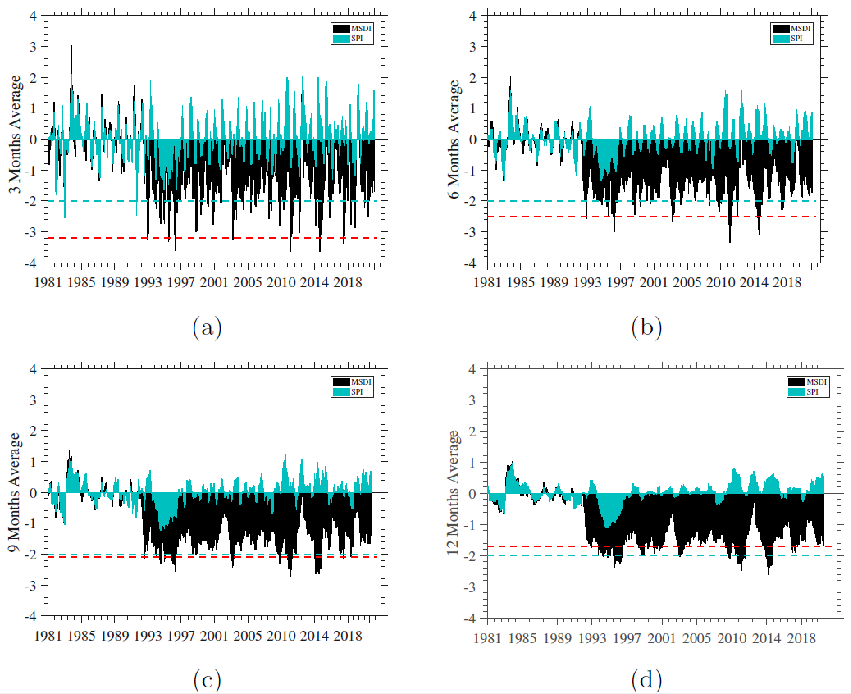}
	\caption{The MSDI and SPI for 3, 6, 9, 12 months for the period of 1981-2021 are shown respectively in (a), (b), s(c) and (d). The highest downward peaks represent years of the occurrence of severe droughts. The red and green dash lines represent threshold limits of extremely dry conditions of MSDI and SPI respectively.} 
	\label{MSDI_spi_panel}
\end{figure}

In this section, we discuss the applicability of the new metric developed in identifying the severity of the paddy-related drought conditions and to compare them against SPI thresholds. The new metric, MSDI is depicted in Fig. \ref{MSDI_panel} whereas the SPI is depicted in Fig. \ref{MSDI_spi_panel}. Fitting this long-term precipitation record to a gamma probability distribution is the first step. Once the gamma probability distribution is determined, the cumulative probability of observed precipitation is computed and then inverse transformed by a standard normal distribution with mean $0$ and variance $1$. The resulting index is the SPI \cite{mckee1993relationship}.
In Fig. \ref{MSDI_panel}, if an MSDI value extend beyond red (bold) line, it is indicative of an extreme dry event whereas if the index value falls between the red (thin) line and red (bold) line it is categorized as a dry condition. The rest can be categorized under the normal/healthy conditions (see Table \ref{tab:tresh}).

Sri Lanka experienced a widespread drought in 1992 with over 18 drought incidents recorded \cite{ref7}. According to Fig. \ref{MSDI_panel}, this drought is shown as extremely dry in 3, 6, 9 and 12 months time scales but not very significant in 9 or 12 months. This confirms that the drought of 1992 was characterized by numerous minor drought events recorded in sub-intervals, rather than a single continuous drought episode. However, the SPI failed to identify this drought episode in its 6, 9, or 12 months SPI charts in Fig. \ref{MSDI_spi_panel}. The CBSL annual report in \cite{cbslreport} states that the paddy production is declined in the cultivation year 1982 due to adverse weather conditions. Additionally, it states that it's worth noting that there is a significant 13\% drop in paddy production during the Yala season of 1983 when compared to the previous Yala season. As per this report, this notable decline in paddy production during Yala $1983$ shows the abandoned or suffering crop damage due to the drought conditions that prevailed during the first half of the year \cite{cbslreport}. Furthermore, based on another historical record, the period from 1982-1983 is categorized as a dry period \cite{ref7}. 
Our developed MSDI, computed for 3, 6, 9, and 12 months, consistently reflects these dry conditions for $1982-1983$ period, as it consistently hovers above the threshold of extreme dryness (represented by the red (bold) line) and below the dry limit (red (thin) line). The length of the peaks of $1983$ is longer than $1982$ as in Fig. \ref{MSDI_panel}, and it aligns with the statement of the CBSL annual report-1983. That is, compared to $1982$ Yala season, there's a higher paddy loss in $1983$. Comparing the SPI, in 3 months it identifies $1983$ as an extreme drought situation while $1982$ as a dry situation. Furthermore, in $6,9$ timeframes SPI only denoted $1983$ as a dry situation. In long run, SPI fails in identifying the dry scenarios for both timeframes as indicated in Fig. \ref{MSDI_spi_panel}.

Referring the historical recordings of CBSL annual report -1987, it emphasizes that the decline of the paddy production in $1987$ was mainly due to the prolonged drought conditions during the cultivation year 1986-1987 \cite{cbslreport}. Further, it highlights that this was the second consecutive year in which paddy production dropped where the estimated 1.39 million metric tons was a substantially declined by 18\%.  If we consider the developed MSDI (given in Fig. \ref{MSDI_panel}), in long run basically for 12 months period it clearly depicts a dry scenario for the period of 1986-1987 where as SPI shows a completely normal situation in long run as its peaks are above its standard thresholds. In both 6, 9 months cases MSDI was successful in identifying 1987 as a drought condition which proves the long term existence of the drought in this cultivation period. SPI is only succeeded in identifying the drought scenario in the short term that is in 3 months case. Over the long run, the SPI consistently falls outside its standard threshold limits, indicating its ineffectiveness in identifying drought situations in prolong period as indicated in Fig. \ref{MSDI_spi_panel}.

Due to the drought condition prevailed in the 1988-1989 cultivation era, the paddy production has been significantly reduced according to the CBSL-1989 report \cite{cbslreport}. Moreover, it highlights that, fall in paddy production was reflected in both 1988-1989 Maha and 1989 Yala, seasons. Considering our MSDI for 9 months period it identifies a dry period for the cultivation year 1988-1989, but SPI is not capable in identifying the drought scenario in this timeframe as its peaks lies outside its standard threshold limit, but SPI identify this drought event in 3 months as shown in Fig. \ref{MSDI_spi_panel}.  

For the period of 1995-1996 CBSL annual report -1996 in \cite{cbslreport} states that the paddy production in Anuradhapura district dropped by 61\% from 190,000 metric tons produced during Maha 1994-1995 to 74,000 metric tons in Maha 1995-1996 due to the occurred drought situation. Referring the MSDI for the cultivation year 1996 it is clearly visible that all the peaks lie below the extremely dry situation for 3, 6, 9, 12 timeframes. This depicts the existence of this drought event for a prolonged period.This suggests that the drought's severity is significantly greater. This observation aligns with the CBSL report, which notes that paddy production, reaching its peak in 1995, experienced a sharp decline of 27\% to reach 2.1 million metric tons—the lowest recorded level since 1979 \cite{cbslreport}. However, SPI categorizes this drought occurrence as a dry scenario. Consequently, when comparing the two indices, MSDI emerges as the more dependable tool for recognizing this drought situation in accordance with historical records. 

According to the CBSL-2001 report, the extents sown and harvested paddy during both Maha, Yala seasons dropped due to adverse weather conditions that prevailed during the year  \cite{cbslreport}. The MSDI calculations for the 3, 6, and 9 months periods effectively pinpoint a dry condition, and over an extended duration, the MSDI surpasses the extremely dry threshold, indicating the persistence of this drought condition for a prolonged period (see Fig. \ref{MSDI_panel}). However, it's worth noting that SPI does not accurately represent this dry condition for this cultivation year as shown in Fig. \ref{MSDI_spi_panel}. Similarly in the cultivation year 2004, the MSDI exceeds the extremely dry threshold for all timeframes. However this can be proven using the historical record, CBSL-2004 annual report \cite{cbslreport}. The report indicates that the agriculture sector's contribution to the GDP declined from 19\% in 2003 to 18\% in 2004, primarily due to reduced paddy production. This decline is attributed to drought conditions in the Anuradhapura and Kurunegala districts during the Maha season, resulting from the failure of the Northeast monsoon and the delayed onset of the Southwest monsoon in the Yala season  \cite{cbslreport}. 

For year 2014, the CBSL 2014 annual report in \cite{cbslreport} highlights a slowdown in the agriculture sector, with a growth rate of only 0.3\% compared to 4.7\% in the previous year. This deceleration is primarily attributed to adverse weather conditions that prevailed throughout the year. Notably, while the SPI have not indicated this adverse drought scenario accurately, the MSDI effectively denotes peaks for drought conditions for 3, 6, 9, and 12 months periods. Additionally, as documented in historical records, this drought condition persisted for an entire year, as evidenced by the visible peaks even in the 12 months MSDI panel (Fig. \ref{MSDI_panel}). During the drought period spanning 2016-2017, the MSDI peaks fell within the range between the extremely dry and dry threshold limits. The CBSL 2016 annual report in \cite{cbslreport} highlights the significant impact of adverse weather conditions, particularly the prolonged drought, on agricultural performance throughout that year. Interestingly, while the MSDI accurately identifies this drought situation (see Fig. \ref{MSDI_panel}), the SPI, in contrast, does not recognize it and instead categorizes it as a normal scenario (see Fig. \ref{MSDI_spi_panel}). 
All in all, the calculated two-dimensional MSDI incorporating rainfall with temperature outperforms the SPI when calibrated against historical drought records.



\newpage
\section{Conclusions}

As outlined in the introduction, droughts result from two primary factors: (1) prolonged periods of elevated temperatures and (2) extreme precipitation events. Consequently, our drought severity assessment system takes into account the effects of these weather variables over extended time periods. The development of this two-dimensional index system involved calibration using historical drought events, demonstrating that the index aligns with the severity experienced during those drought occurrences.\\
The North Central (NC) region's semi-arid climate makes it highly susceptible to droughts, which pose significant challenges to the agricultural sector. Droughts occur during prolonged periods of reduced precipitation, aggravated by elevated temperatures. As a result, the paddy cultivation industry in this area is heavily affected. NC has a history of experiencing numerous droughts, including instances in 1935-1937, 1947-1949, 1953-1956, 1965, 1974-1977, 1982-1983, 1986-1987, 1992,  1995-1996, 2001, 2004, 2013-2014, and 2016-2017 \cite{ref9,ref44,ref47}.
In this study, we employ a weather data modeling approach to assess drought severity, utilizing the FGM copula due to its capacity to capture extreme weather events through tail dependencies. Our analysis focuses on two key factors contributing to droughts, namely rainfall and temperature, over a 40-year period spanning from 1981 to 2021. The marginal distributions derived for rainfall and temperature are represented respectively by the beta distribution (Fig. \ref{fig:marginals}) and a Gaussian mixed model (Fig. \ref{fig:marginals}). Following the fitting of the copula to the transformed data, we compute the MSDI. These MSDI values serve as indicators of drought severity, and we establish thresholds for different levels of seventies, including normal, dry, and extremely dry conditions (see Table \ref{tab:tresh}).\\
Furthermore, the developed two-dimensional index system has been compared with the already existing, most widely used drought monitoring index SPI \cite{ref30}. It's noteworthy that the SPI is built upon the analysis of precipitation data only \cite{ref48}. The common approach of developing the SPI involves fitting a gamma distribution to the precipitation data \cite{ref49}. However, in our study we integrate both precipitation and ground temperature to calculate MSDI. While constructing the SPI, we followed the typical procedure, that is, fitting a gamma distribution to the rainfall data though it best fits for a beta distribution as detailed earlier. Though, for MSDI we employed a beta distribution for rainfall according to the fitting choice of AIC value. Afterwards, evaluating the developed MSDI against SPI, the historical data played a crucial role in comparing and decision-making. Upon comparing these two indices MSDI consistently outperforms SPI. Furthermore, we can conclude that MSDI works more accurately for longer time frames.\\
In conclusion, we emphasize the significance of incorporating temperature data into the index calculation, as opposed to relying solely on rainfall, as done in the SPI. We illustrate the superior performance of MSDI compared to SPI. Furthermore, we underscore the importance of accurately characterizing the marginal distributions of weather variables. Our MSDI system's validity is affirmed through its successful validation against historical drought events. When contrasting with existing drought severity evaluations found in the literature, these assessments typically employ drought variables like drought duration and drought intensity to construct frequency distributions. For prospective research endeavors, the novel MSDI can be employed to compute these drought variables, forming frequency distributions for drought occurrences. This approach allows for a comparison of accuracy against conventional models.


\begin{thebibliography}{35}
\providecommand{\natexlab}[1]{#1}
\providecommand{\url}[1]{{#1}}
\providecommand{\urlprefix}{URL }
\providecommand{\doi}[1]{\url{https://doi.org/#1}}
\providecommand{\eprint}[2][]{\url{#2}}
 \bibcommenthead

\bibitem[{Abeysingha and Rajapaksha(2020)}]{ref47}
Abeysingha N, Rajapaksha U (2020) Spi-based spatiotemporal drought over sri
  lanka. Advances in Meteorology 2020:10

\bibitem[{Akaike(1974)}]{ref25}
Akaike H (1974) A new look at the statistical model identification. IEEE
  Transactions on Automatic Control 19(6):716--723

\bibitem[{CBSL(2021{\natexlab{a}})}]{cbslreport}
CBSL (2021{\natexlab{a}}) Cbsl annual report. Retrieved at:
  \url{https://www.cbsl.gov.lk/en/publications/economic-and-financial-reports/annual-reports}

\bibitem[{CBSL(2021{\natexlab{b}})}]{ref5}
CBSL (2021{\natexlab{b}}) Central bank of sri lanka annual report 2021.
  Retrieved at:
  \url{https://www.cbsl.gov.lk/en/publications/economic-and-financial-reports/annual-reports/annual-report-2021}

\bibitem[{Center(2021)}]{NASApower}
Center NLR (2021) Nasa power data access viewer. Retrieved at:
  \url{https://power.larc.nasa.gov/data-access-viewer/}

\bibitem[{Chandrasiri et~al(2020)Chandrasiri, Galagedara, and Mowjood}]{ref10}
Chandrasiri S, Galagedara L, Mowjood M (2020) Impacts of rainfall variability
  on paddy production: A case from bayawa minor irrigation tank in sri lanka.
  Paddy and Water Environment 18:443--454

\bibitem[{DCS(2023)}]{ref21}
DCS (2023) Statistical pocket handbook: Department of census and statistics.
  Retrieved at: \url{http://www.statistics.gov.lk/ref/PocketBook2022_Si_Ta}

\bibitem[{DMC(2017)}]{ref11}
DMC (2017) Disaster information management system-sri lanka. Retrieved at:
  \url{http://www.desinventar.lk/}

\bibitem[{EOS(2023)}]{ref22}
EOS (2023) Growing rice: Sowing, cultivating, and harvesting. Retrieved at:
  \url{https://eos.com/blog/how-to-grow-rice}

\bibitem[{Fukagawa and Ziska(2019)}]{fukagawa2019rice}
Fukagawa NK, Ziska LH (2019) Rice: Importance for global nutrition. Journal of
  nutritional science and vitaminology 65(Supplement):S2--S3

\bibitem[{Ganguli and Reddy(2012)}]{ref32}
Ganguli P, Reddy M (2012) Risk assessment of droughts in gujarat using
  bivariate copulas. Water resources management 26:3301--3327

\bibitem[{Genest et~al(2009)Genest, R{\'e}millard, and Beaudoin}]{ref28}
Genest C, R{\'e}millard B, Beaudoin D (2009) Goodness-of-fit tests for copulas:
  A review and a power study. Insurance: Mathematics and Economics 44:119--213

\bibitem[{Gunawardhana and Dharmasiri(2016)}]{ref9}
Gunawardhana L, Dharmasiri L (2016) Drought hazard and managing its impacts
  through the disaster management approach: A study in the north central
  province of sri lanka. Paper presented at the International Research
  Symposium Rajarata University of Sri Lanka

\bibitem[{Hao and AghaKouchak(2014)}]{ref30}
Hao Z, AghaKouchak A (2014) Multivariate standardized drought index: a
  parametric multi-index model. Advances in Water Resources 57:12--18

\bibitem[{Haugh(2016)}]{ref34}
Haugh M (2016) An introduction to copulas. IEOR E4602: quantitative risk
  management Lecture notes Columbia University

\bibitem[{Khan et~al(2021)Khan, Faisal, Hashmi, Nazeer, Ali, and
  Hussain}]{ref42}
Khan MA, Faisal M, Hashmi M, et~al (2021) Modeling drought duration and
  severity using two-dimensional copula. Journal of Atmospheric and
  Solar-Terrestrial Physics 214:105530

\bibitem[{Lee et~al(2013)Lee, Modarres, and Ouarda}]{ref41}
Lee T, Modarres R, Ouarda T (2013) Data-based analysis of bivariate copula tail
  dependence for drought duration and severity. Hydrological Processes
  27(10):1454--1463

\bibitem[{Manesha et~al(2015)Manesha, Vimukthini, and Premalal}]{ref13}
Manesha S, Vimukthini S, Premalal K (2015) Develop drought monitoring in sri
  lanka using standard precipitation index (spi). Sri Lanka J Meteorol 1:64--71

\bibitem[{McKee et~al(1993{\natexlab{a}})McKee, Doesken, and Kleist}]{ref49}
McKee T, Doesken N, Kleist J (1993{\natexlab{a}}) The relationship of drought
  frequency and duration to time scales. Paper presented at the Eighth
  Conference on Applied Climatology, 17-22 January 1993, Anaheim, California

\bibitem[{McKee et~al(1993{\natexlab{b}})McKee, Doesken, Kleist
  et~al}]{mckee1993relationship}
McKee TB, Doesken NJ, Kleist J, et~al (1993{\natexlab{b}}) The relationship of
  drought frequency and duration to time scales. In: Proceedings of the 8th
  Conference on Applied Climatology, California, pp 179--183

\bibitem[{MoE(2021)}]{ref12}
MoE (2021) Ministry of environment-sri lanka. Retrieved at:
  \url{https://env.gov.lk/web/index.php/en/}

\bibitem[{Nelsen(2006)}]{ref24}
Nelsen RB (ed)  (2006) An introduction to copulas. Springer Science \& Business
  Media, New {Y}ork

\bibitem[{Nianthi(2019)}]{ref14}
Nianthi R (2019) Farmers’ responses to drought: Dry zone of sri lanka:(case
  study in medirigiriya). Case Studies Journal 5

\bibitem[{Otkur et~al(2021)Otkur, Wu, Zheng, Kim, and Lee}]{otkur2021copula}
Otkur A, Wu D, Zheng Y, et~al (2021) Copula-based drought monitoring and
  assessment according to zonal and meridional temperature gradients.
  Atmosphere 12(8):1066

\bibitem[{Pandey et~al(2018)Pandey, Das, Jhajharia, and Pandey}]{ref27}
Pandey P, Das L, Jhajharia D, et~al (2018) Modelling of interdependence between
  rainfall and temperature using copula. Modeling Earth Systems and Environment
  4:867--879

\bibitem[{Papadopoulos et~al(2021)Papadopoulos, Spiliotis, Gkiougkis, Pliakas,
  and Papadopoulos}]{ref43}
Papadopoulos C, Spiliotis M, Gkiougkis I, et~al (2021) Fuzzy linear regression
  analysis for groundwater response to meteorological drought in the aquifer
  system of xanthi plain, ne greece. Journal of Hydroinformatics
  23(5):1112--1129

\bibitem[{Reddy and Ganguli(2012)}]{ref17}
Reddy M, Ganguli P (2012) Bivariate flood frequency analysis of upper godavari
  river flows using archimedean copulas. Water Resources Management
  26(14):3995--4018

\bibitem[{Reyes et~al(2022)Reyes, Rangel, and Herazo}]{ref48}
Reyes L, Rangel H, Herazo L (2022) Adjustment of the standardized precipitation
  index (spi) for the evaluation of drought in the arroyo pechel{\'\i}n basin,
  colombia, under zero monthly precipitation conditions. Atmosphere 13(2):236

\bibitem[{Senanayake and Premaratne(2016)}]{ref6}
Senanayake S, Premaratne S (2016) An analysis of the paddy/rice value chains in
  sri lanka. Asia-Pacific Journal of Rural Development 26(1):105--126

\bibitem[{Senatilleke et~al(2023)Senatilleke, Sirisena, Gunathilake, Muttil,
  and Rathnayake}]{ref44}
Senatilleke U, Sirisena J, Gunathilake M, et~al (2023) Monitoring the
  meteorological and hydrological droughts in the largest river basin (mahaweli
  river) in sri lanka. Climate 11(3):57

\bibitem[{Seyedabadi et~al(2020)Seyedabadi, Kavianpour, and
  Moazami}]{seyedabadi2020multivariate}
Seyedabadi M, Kavianpour M, Moazami S (2020) Multivariate drought risk analysis
  based on copula functions: a case study. Water Supply 20(6):2375--2388

\bibitem[{Shiau(2006)}]{ref26}
Shiau J (2006) Fitting drought duration and severity with two-dimensional
  copulas. Water resources management 20(795--815):795--815

\bibitem[{Sklar(1959)}]{ref23}
Sklar A (1959) Fonctions de r{'e}partition {`a} n dimensions et leurs marges.
  Publications de l'Institut de Statistique de l'Universit{'e} de Paris
  8:229--231

\bibitem[{Tuong and Bouman(2003)}]{ref7}
Tuong T, Bouman B (2003) Rice production in water-scarce environments. Water
  productivity in agriculture: Limits and opportunities for improvement
  1:13--42

\bibitem[{Wang et~al(2019)Wang, Wang, Yang, Zhao, Zhang, Li, and
  Hussain}]{ref31}
Wang F, Wang Z, Yang H, et~al (2019) Copula-based drought analysis using
  standardized precipitation evapotranspiration index: A case study in the
  yellow river basin, china. Water 11(6):1298

\end{thebibliography}

\end{document}